\documentstyle[aps,twocolumn,pre,psfig,floats]{revtex}     
\draft                                         

\newcommand{\bnabla}{\mbox{\boldmath{$\nabla$}}}
\newcommand{\pder}[2]{\frac{\partial {#1}}{\partial {#2}}}
\newcommand{\der}[2]{\frac{d {#1}}{d {#2}}}
\newcommand{\norm}[1]{\hat{\mbox{\boldmath{#1}}}}

\begin{document} 
 
\twocolumn[\hsize\textwidth\columnwidth\hsize\csname @twocolumnfalse\endcsname

\title{Metastability of a granular surface in a spinning bucket.}

\author{Chuck~Yeung} 
 
\address{ 
Division of Science, Pennsylvania State University at  Erie, 
The Behrend College, Erie, PA 16536 USA \\ 
} 
 
\date{\today} 
 
\maketitle 
 
\widetext 
  
\begin{abstract} 
The surface shape of a spinning bucket of granular material is studied using a
continuum model of surface flow developed by Bouchaud et al.\ and Mehta et al. 
An experimentally observed central subcritical region is reproduced by the
model. The subcritical region occurs when a metastable surface becomes
unstable  via  a {\em nonlinear} instability mechanism.  The nonlinear
instability mechanism destabilizes the surface in large systems while a linear
instability mechanism is relevant for smaller systems. 
The range of angles in which the
granular surface is metastable vanishes with increasing system size. \\
\ \\
to be published in Physical Review E
\end{abstract} 
 
 
\narrowtext ]                   

\section{Introduction}

Granular materials display very complex dynamics due to their both
fluid-like and solid-like characteristics \cite{NEDDERMAN,REVIEWS}. Even
simple experiments can produce unexpected results.  Examples are  recent
experiments in which measurements were taken of the granular surface of 
a bucket of sand spun about its  cylindrical axis \cite{VAVREK,BAXTER}.
At low rotation rates a central region was observed in which the slope is
significantly  less than the critical slope \cite{BAXTER,NOTE,MEDINA}.
This central subcritical region was conjectured to be due to the inertia
of the flowing grains \cite{BAXTER} since it could not be
explained in terms of a simple theory which assumed the surface was
everywhere critical \cite{COULOMB}.

Many models have been developed to describe the evolution of a granular
surface \cite{DUKE90,OUYANG,BOUCHAUD,MEHTA,BARKER,BOUTREUX,MAKSE,HEAD}.  These models
often assume that the flow is restricted to a thin layer of grains near
the surface.  Such models have been used extensively to study avalanches of a
granular material in a horizontal rotating drum \cite{JAEGER89}. However, the
spinning bucket experiment is a better testing ground for these ideas
since, at least at low rotation rates, the fundamental assumption of a
thin flowing layer is more likely to be true \cite{BARKERNOTE}.


The purpose of this paper is two-fold.  The first is to apply the flow
models to the spinning bucket experiment.  The second is to explore the
source and limits of the metastable behavior observed in the model.  In
particular, a continuum description of the surface flow developed by
Bouchaud et al.\ \cite{BOUCHAUD} and Mehta et al.\ \cite{MEHTA} is used. 
This model includes the effect of the inertia of the flowing grains and
is known to exhibit  metastability via a linear instability mechanism
\cite{BOUCHAUD}, i.e., the surface remains metastable for a range of
angles beyond the minimum angle of repose. This difference in the minimum
and maximum stable angles was argued to be the physical source of the
Bagnold angle \cite{BOUCHAUD}.

Here I show that this model qualitatively reproduces the central subcritical
region. I determine the mechanism creating the subcritical region and show that
it is closely intertwined with the metastability of the model. The metastable
surface becomes unstable via a {\em nonlinear} instability, i.e., an
instability to very small but non-infinitesimal perturbations. This nonlinear
instability determines the dynamics for ``large'' systems while the linear
instability is dominant for ``small'' systems.  
The range of metastable slopes and, hence the Bagnold
angle, depends on the system size and vanishes in the limit of large systems.

\section{Summary of Experiment}

\begin{figure}
\ \\

\centerline{
\psfig{figure=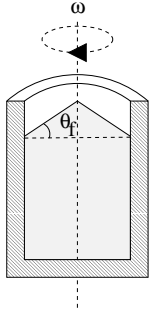,width=0.75in}
}
\ \\

\caption{
A cut-away view of the the spinning bucket
experiment.   A conical pile is prepared at the angle of
repose $\theta_f$.  The cylinder is then spun about its vertical
axis at rotation rate $\omega$ and the resulting surface shape
is measured.
\label{FIG:BUCKET}
}

\end{figure}

Figure \ref{FIG:BUCKET} shows the  setup for the spinning bucket
experiment \cite{VAVREK,BAXTER}.   A bucket of radius $R_o$ is partially
filled with common building sand.  A conical pile is prepared at
the angle of repose, $\theta_f =  34 \pm 1 ^{o}$, by slowly dropping the
plastic spheres onto the center of the stationary bucket. (
Here we label the angle of repose $\theta_f$ since,
in an ideal Coulomb material, the surface is everywhere critical
and the angle of repose is the same as the angle of internal friction
\cite{NEDDERMAN}). The rotation rate $\omega$ is then
slowly increased until it reaches the desired value and the resulting
surface shape is measured.

For an ideal Coulomb material the granular surface shape is found by
assuming that the friction force is saturated and relating the net force
on a surface element  to the centripetal acceleration
\cite{VAVREK,BAXTER,MEDINA}.  The critical slope $S_{-}$ is a function of
radial distance $r$ and $\omega$,
\begin{equation}
	 S^{-}(r) =  \frac{dh}{dr} =
		\frac{ (r/R_o) \Omega^2 - \tan \theta_f }{
			1 + (r/R_o) \Omega^2  \tan \theta_f },
				\label{EQ:CRITICAL}
\end{equation}
where $h(r)$ is the local height,  $\Omega = \sqrt{ \omega^2 R_o / g }$ is the
dimensionless rotation rate, $g$ is the gravitational acceleration and $R_o$ is
the bucket radius. The $-$ superscript indicates that the friction force is
inwards since the grains flow outwards.  Note that $S^{-}(r)$ is negative for
small $\Omega$ so that a unstable surface corresponds to $\partial h/\partial r
< S^{-}(r)$.

\begin{figure}

\centerline{
\psfig{figure=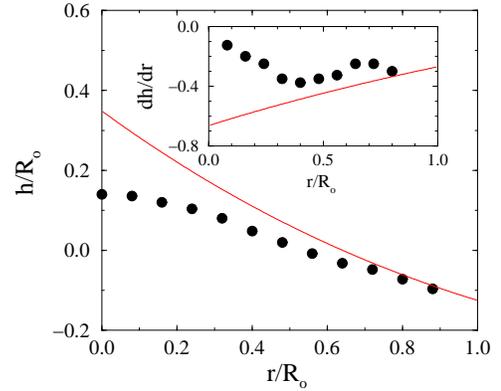,width=2.5in}
}

\caption{
\label{FIG:LOWW}
The surface shape at $\Omega =
\protect\sqrt{\omega^2 R_o/g} = 0.578$ as given in Ref.\ \protect\cite{BAXTER}. 
The center of the bucket is at $r = 0$ and the edge at $r/R_o = 1$.
The dashed line is the  
result for a ideal Coulomb material with $\theta_f = 34^{0}$
and the points are the experimental data.  The inset shows the
experimental and theoretical slopes.
The experiment and theory agrees near the
outer edge of the bucket but 
the slope at the center is significantly less negative than that
of an ideal Coulomb material.
}

\end{figure}

Experimentally, the surface agrees well with Eq.\ [\ref{EQ:CRITICAL}]
for intermediate rotation rates $2 \lesssim \Omega  \lesssim 5$
\cite{VAVREK,BAXTER} but less well for larger and smaller $\Omega$.
This paper will focus on the low rotation rates, $\Omega
\lesssim 2$. Figure \ref{FIG:LOWW} shows the experimental surface
shape  obtained by Baxter for $\Omega = 0.578$ along with the
prediction for a ideal Coulomb material using $\theta_f = 34^{0}$
\cite{BAXTER}. The corresponding slopes are shown in the inset. There
is reasonable agreement between experiment and theory at the outer
edge ($r \gtrsim R_o/2$). However, in the center, $r \lesssim R_o/2$,
the slope is much less than the critical slope. This was thought to be
because the critical theory neglects the effects of grain inertia 
\cite{BAXTER}.

\section{The model}

To analyze the spinning bucket experiment we use a model of
the granular surface evolution developed by Bouchaud et
al.\  \cite{BOUCHAUD} and Mehta et al.\ \cite{MEHTA}.  The
model assumes that all grains are stationary except for a thin
layer of flowing grains on the top.  We define $h( {\bf r},
t)$ as the local height of the immobile pile and $\rho( {\bf
r}, t)$ as the thickness of the rolling layer.  Here ${\bf r}$ is the
two-dimensional vector giving
the projection of the local position onto the $x-y$ plane.

The dynamics of the rolling layer $\rho({\bf r},t)$ is given by
\begin{equation}
	\frac{ \partial \rho }{ \partial t }
		=
	-\bnabla \cdot ( {\bf v} \rho ) +
	D_{o} \nabla^2 \rho +
	\Gamma_{o}( \{ h \}, \{ \rho \} ),
		\label{EQ:RHO0}
\end{equation}
where ${\bf v} = v_o \norm{r}$ is the local velocity of the
rolling layer, $D_o$ is the diffusion constant and
$\Gamma_{o}$ is the rate of conversion from immobile to
mobile grains and $\bnabla$ and $\nabla^2$ are
the two-dimensional gradient and laplacian respectively.
Assuming that the grains flow outward, the
conversion rate is:
\begin{eqnarray}
	\Gamma_{o}( \{ h \}, \{ \rho \} )
	& = & - \rho
	\left[ \gamma_o \left( \pder{h}{r} - S^{-}(r) \right)
	+ \kappa_{o}  \nabla^2 h  \right]
		\nonumber \\
	& & - \alpha_{o} \left[ \pder{ h}{r} - S^{-}(r) \right]
		{\cal H}\left( \pder{ h}{r} - S^{-}(r) \right),
	\label{EQ:GAMMA0}
\end{eqnarray}
where ${\cal H}$ is the Heaviside function. The first term
in  $\Gamma_o$ accounts for the conversion of static grains
to rolling grains when the slope is steeper than the
critical slope ($\partial h/\partial r < S^{-}$) and vice
versa for a shallower slope. The  $\rho \kappa_o \nabla^2 h$
term  causes valleys to fill and peaks to smooth.   The last
term in $\Gamma_o$ accounts for  the jarring loose of static
grains by the rotation of the bucket even in the absence of
a flowing layer. Finally, assuming that bulk rearrangement
of material can be neglected, conservation of total material
requires
\begin{equation}
	\pder{h}{t} = - \Gamma_o( \{ h \}, \{ \rho \} ).
		\label{EQ:H0}
\end{equation}
For $\alpha_o = 0$ the model is the same as that introduced
by Bouchaud et al.\ \cite{BOUCHAUD}.  Mehta et al.\ allowed
for a nonzero $\alpha_o$ and also included an additional
bulk arrangement term  \cite{MEHTA}.

To compare with experiment we require a rough estimate
of the model parameters:

\begin{itemize}

\item
The speed  of the rolling grains $v_o$ can be estimated from
the speed of a grain falling through a grain
diameter $d_o$: $v_o \approx \sqrt{ 2 g d_o }$.

\item
A narrow bump of rolling grains starting at the center will
reach the edge of the bucket in time $t_o = R_o/v_o$.
Diffusion causes this bump to spread to a width $\Delta x =
\sqrt{D_o t_o} = \sqrt{D_o R_o/v_o}$ when it reaches the
edge. Since $\Delta x \sim \sqrt{R_o}$ the width of the bump
is much less than the system size for large $R_o$.
Therefore the ``large'' system limit corresponds to
\begin{equation}
	\frac{ (\Delta x)^{2} }{R_o^2} =
		\frac{ D_o }{ v_o R_o } \ll 1.
				\label{EQ:SMALL}
\end{equation} 
The ``small'' limit corresponds to the width of the bump
being of the same order as the system or $D_o / ( v_o R )
\gtrsim 1$ \cite{PECLET}.  

\item $\gamma_o ~\Delta S~ d_o/v_o$ is the probability that
a rolling grain jars loose a static grain as it rolls over
it. A typical excess slope of $\Delta S = 0.1$ and,
assuming a 10 \% probability that  the static grain is
converted to rolling, gives $\gamma_o \approx ~v_o/d_o$.

\item The ratio $\kappa_o/\gamma_o$ defines a length scale
on which the hole filling/peak smoothing mechanism
dominates.  This is   important only at very small length
scales on the order of a few grain diameters so $\kappa_o
\approx d_o \gamma_o$.

\item $\alpha_o$ depends on the noise in the apparatus.  The
generation of rolling material by noise is assumed to be
much less effective than the convection so that $\alpha_o
\ll v_o$.

\end{itemize}

Equations (\ref{EQ:RHO0}-\ref{EQ:H0}) can be rewritten in
dimensionless form by measuring lengths in terms of the
bucket radius $R_o$ and time in terms of $t_o = R_o/v_o$,
\begin{equation}
	\pder{\rho }{ t }
		=
	-\bnabla \cdot (\norm{r} \rho) +
	D \nabla^2 \rho +
	\Gamma( \{ h \}, \{ \rho \} ),
		\label{EQ:RHO}
\end{equation}
where
\begin{eqnarray}
	\Gamma( \{ h \}, \{ \rho \} )
	& = & - \rho
	\left[ \gamma \left( \pder{h}{r} - S^{-}(r) \right)
	+ \kappa  \nabla^2 h  \right]
		\nonumber \\
	& & - \alpha \left[ \pder{ h}{r} - S^{-}(r) \right]
		{\cal H}\left( \pder{ h}{r} - S^{-}(r) \right),
	\label{EQ:GAMMA}
\end{eqnarray}
and
\begin{eqnarray}
	\pder{ h }{t} = - \Gamma( \{ h \}, \{ \rho \} ).
		\label{EQ:H}
\end{eqnarray}
The dimensionless parameters are $D = D_o/(v_o R)$, $\gamma = \gamma_o
R/v_o$, $\kappa = \kappa_o/v_o$ and  $\alpha = \alpha_o/v_o$.

Assuming a large system,
order of magnitude estimates of 
the dimensionless parameters 
can be obtained from the earlier estimates:
\begin{eqnarray}
	D \equiv \frac{ D_o }{ v_o R }   \ll  1,
		& ~~ &
	\gamma \equiv \frac{ \gamma_o R }{ v_o }
		 \approx   \frac{R}{ d_o },
			\nonumber \\
	\kappa \equiv \frac{ \kappa_o }{ v_o }
		\approx  \frac{d_o}{R}~ \gamma,
			& & 
	\alpha \equiv \frac{ \alpha_o }{ v_o }  \ll  1.
\end{eqnarray}
The speed of the rolling layer is unity under this
rescaling. As shown in the next section, the behavior of the
model is most sensitive to $\gamma$ and less sensitive to
the exact value of $D$ and $\kappa$ as long as  $D \ll 1$
and $\kappa/\gamma \ll 1$.

An especially interesting feature of this model is that it displays
metastability and hysteresis.
Bouchaud et al.\ performed a linear stability analysis by balancing the rate at which
a small bump of the rolling layer is convected downhill with the rate at which
the bump is amplifies and diffuses \cite{BOUCHAUD}. They found that the bump
affects the behavior uphill only if the slope exceeds the critical slope by an
amount larger than
\begin{equation}
	\Delta S = S^{-} - \frac{\partial h}{\partial r}
		 > \frac{ v_o^2}{ D_o \gamma_o }
		= \frac{1}{ D \gamma}.
		\label{EQ:LINEAR}
\end{equation} 
The metastable behavior was interpreted as the physical source of the Bagnold angle,
i.e., the excess angle beyond the angle of repose at which a static sandpile
first becomes unstable.  This dynamical
interpretation of the Bagnold angle is 
very different from the usual mechanical interpretation of this angle
\cite{BOUCHAUD}.

\begin{figure}

\centerline{
\psfig{figure=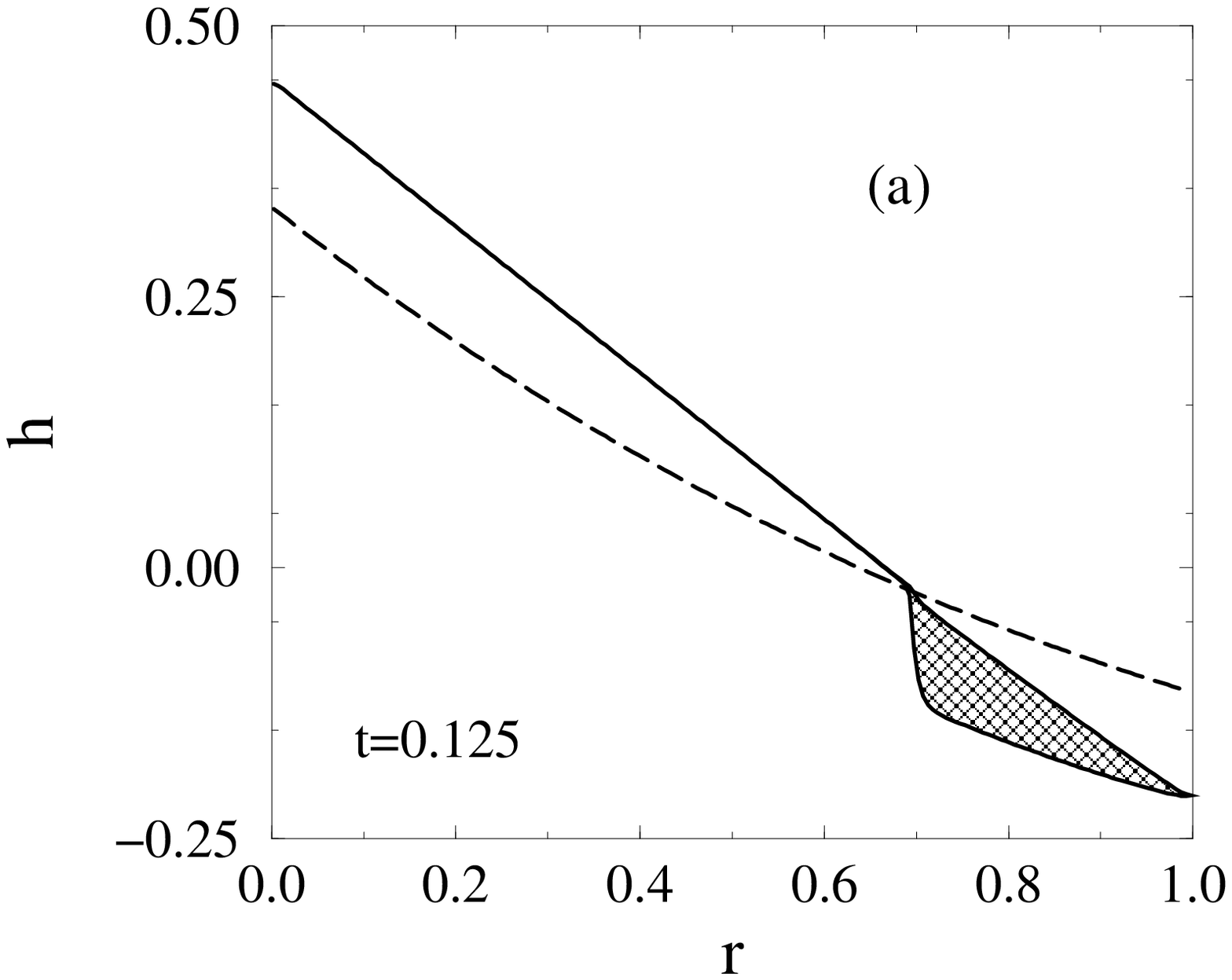,width=2.25in}
}

\centerline{
\psfig{figure=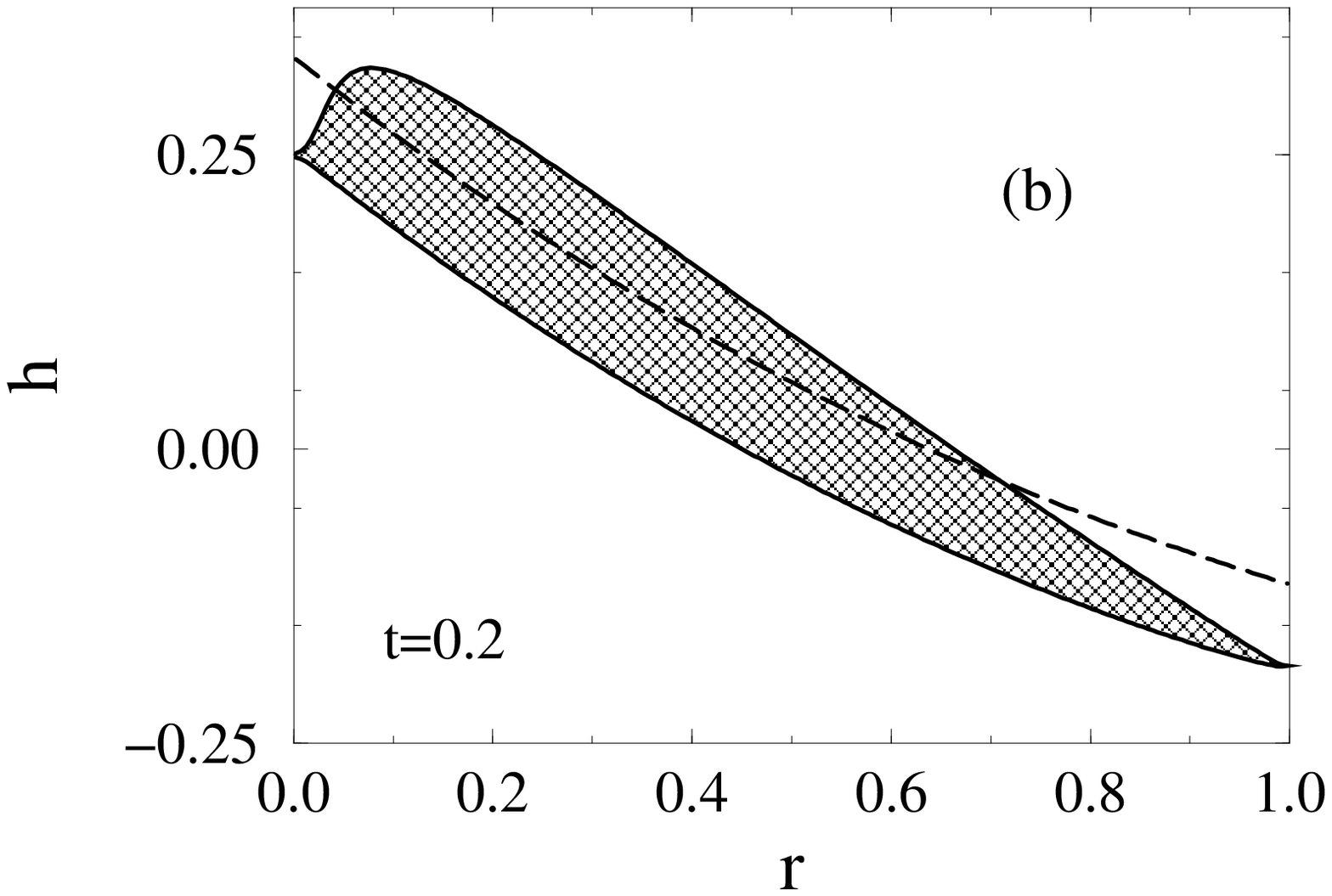,width=2.25in}
}

\centerline{
\psfig{figure=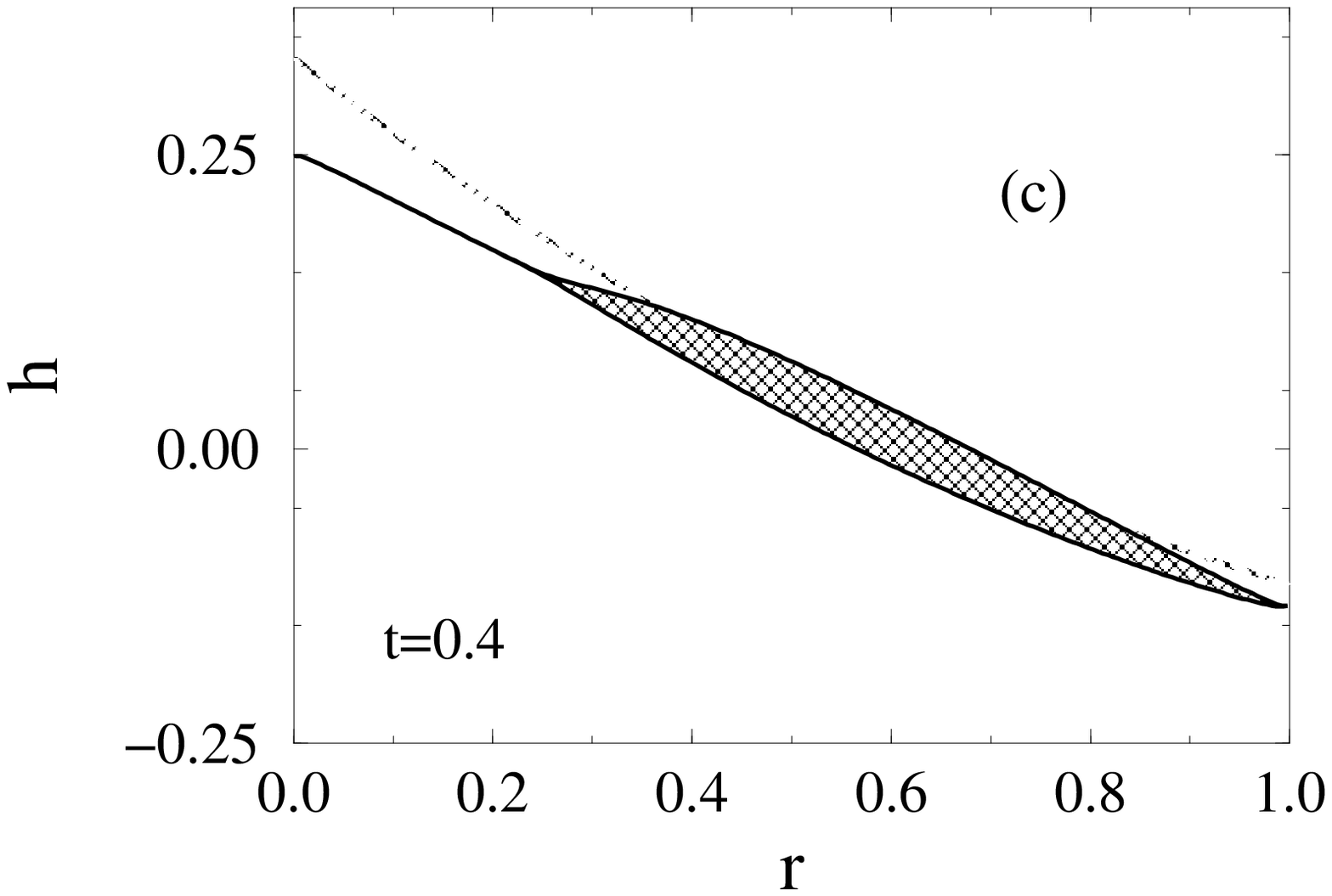,width=2.25in}
}

\centerline{
\psfig{figure=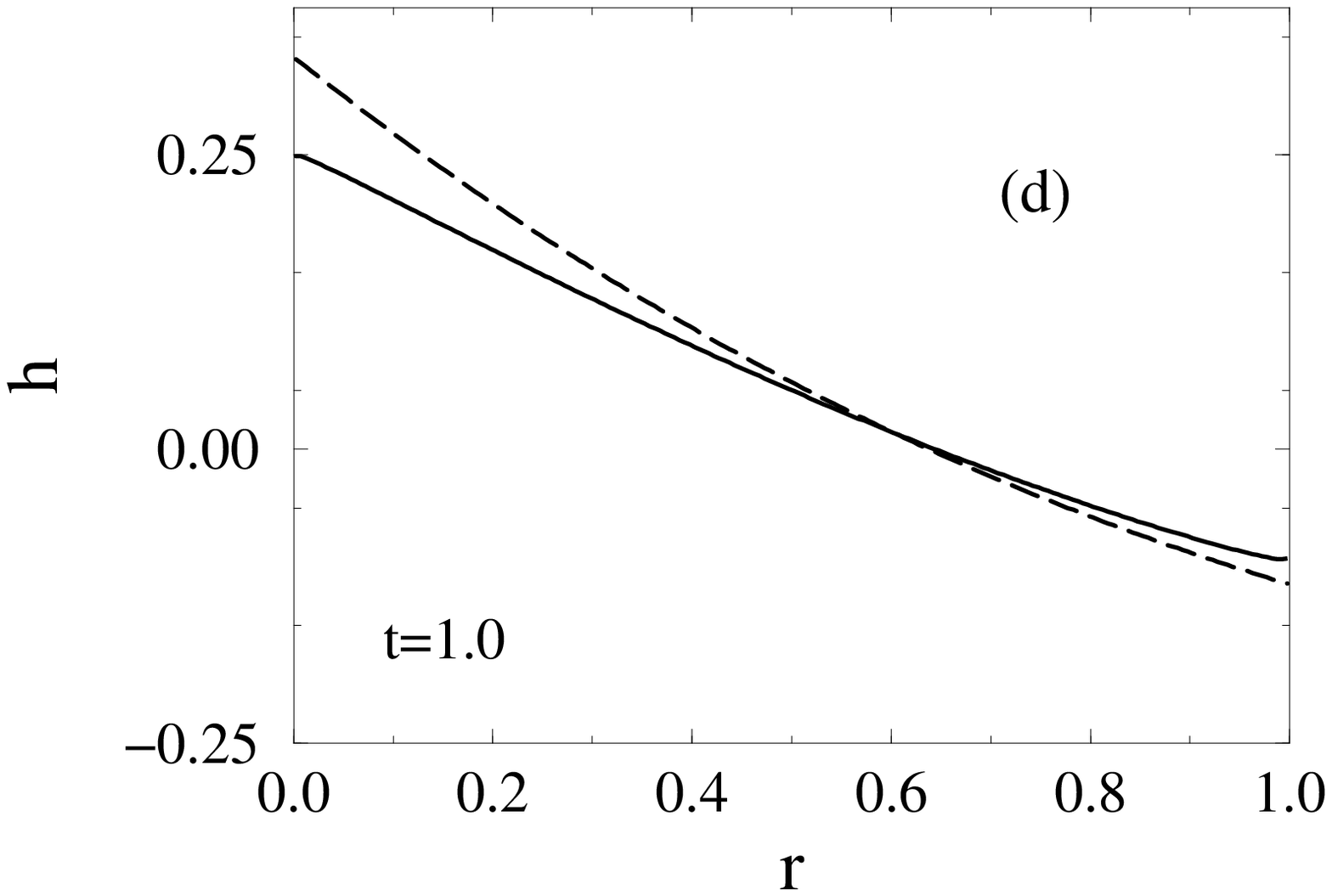,width=2.25in}
}

\caption{
\label{FIG:EVOL}
The evolution of a conical pile at $\Omega = 0.6$.  The dashed line is the
critical surface and the shaded area is the flowing layer.  In (a) a flowing
region  has formed at the edge of the bucket.  This failure zone propagates
uphill reaching the center in (b). The static surface under the flowing layer
closely follows the critical surface. The flowing layer convects downhill (c)
and the final surface (d) has a central region with a slope less than the
critical slope.
}

\end{figure}

\section{Numerical Results}

\subsection{Spinning Bucket}

To analyze the spinning bucket experiment, Eq.\
(\ref{EQ:RHO}-\ref{EQ:H}) were numerically integrated in
polar coordinates with azimuthal symmetry.   Using the
experimental parameters, $\theta_f = 34^{o}$, $d_o
\approx 0.25~mm$ and  $R_o = 12.5~cm$ \cite{BAXTER}, we
obtain $\gamma = R_o/d_o = 500$ and $\kappa/\gamma = 
d_o/R_o = 1/200$.  Assuming a ``large'' system and small
vibrations, we choose $D = 0.001 \ll 1$ and $\alpha =
10^{-8}$. A third-order
Adam-Bashford method was used \cite{NUMERICAL} with mesh-size
$\delta x = 0.002$ and time-step $\delta t = 5 \times
10^{-7}$. Zero slope boundary conditions are applied at $r =
0$ and zero curvature boundary conditions at $r = 1$. 
Any rolling material
reaching the edge is converted to static.  

An initial conical pile is prepared at the critical angle
and the rotation rate $\Omega$ is increased instantaneously
from $\Omega = 0$ to $\Omega = 0.6$.  Since the critical
slope decreases with $\Omega$ the initial surface is
unstable and, as shown in Fig.\ \ref{FIG:EVOL}, an avalanche
is generated. At $t \approx 0.125$  (Fig.\ \ref{FIG:EVOL}a)
a buildup of flowing material is visible at the edge of the
bucket.  This buildup generates a steep local slope in the
static pile which causes the static material uphill to
fail.  The failure then propagates uphill reaching the
center of the bucket at $t \approx 0.2$ (Fig.\
\ref{FIG:EVOL}b).  The static pile beneath the flowing layer
approximately follows the critical curve. The flowing layer
then convects to the edge of bucket (Fig.\ \ref{FIG:EVOL}c)
leaving behind a central subcritical region in which in
which the slope is much less than the critical slope (Fig.\
\ref{FIG:EVOL}d). 

The integration was repeated with different values of $D$, $\gamma$ and
$\kappa$  to
test the robustness of the results.  The central subcritical
region is qualitatively the same for all $\gamma \gtrsim
400$ and $D \lesssim 0.005$ becoming more pronounced for
larger $\gamma$ and smaller $D$.  For  smaller $\gamma$ and
larger $D$ ($\gamma \lesssim 200$ and $D \gtrsim 0.02$) the
final surface follows closely the critical surface. Varying
the ratio $\kappa/\gamma$ or $\alpha$ did not have
noticeable effect as long as they were small but nonzero.

A comparison of model results in Fig.\ \ref{FIG:EVOL} with
the experimental data in Fig.\ \ref{FIG:LOWW} shows that the
essential features of the experiment are reproduced. In
particular, there is a subcritical region in the center and
a region near the edge which follows the critical surface.
However, a more detailed analysis shows two important
differences.  First, the flowing layer is much larger
than the few grains assumed in the model.  Second,
in the experiment, the rotation rate is slowly ramped
up from zero rather than changed instantaneously. Hence
the numerical results corresponds to the actual
experiment only if the final surface is independent
of the initial state, or, if the initial conical surface is
metastable up to an $\Omega$ close to $0.6$.

To mimic the experiment more closely we repeat the
integration while slowly ramping the rotation rate from
$\Omega = 0$ to $\Omega = 0.6$. We monitor the excess slope
before an avalanche occurs, i.e., the amount the slope
of the metastable surface exceed the critical slope, 
and the magnitude of the subcritical region
after an avalanche, i.e., the amount the slope of static
surface falls below the critical slope.

The conical surface is metastable as
$\Omega$ is increased from zero
until the first avalanche occurs at $\Omega \approx 0.2$.
The dynamics of this avalanche is very
similar to that for the instantaneous jump to $\Omega = 0.6$.
The main differences are that the excess slope
before, and the magnitude of the static slope
after the avalanche are both much smaller for the
instantaneous jump to $\Omega = 0.6$. As $\Omega$ is increased
further the critical slope decreases and becomes shallower
than the new static slope.  Eventually a second avalanche
occurs at $\Omega  \approx 0.35$. The process repeats,
leading to  a third avalanche at $\Omega \approx 0.45$ and a
fourth at $\Omega \approx 0.55$. After each avalanche a
subcritical region is observed with  a much smaller
magnitude than for the instantaneous jump. This indicates
that the magnitude of the subcritical region after an
avalanche increases with the excess slope before the
avalanche.  This was confirmed by instantaneously changing
the rotation rate to different values of $\Omega$ and
observing the subcritical region
after the  resulting avalanches.

\begin{figure}

\centerline{
\psfig{figure=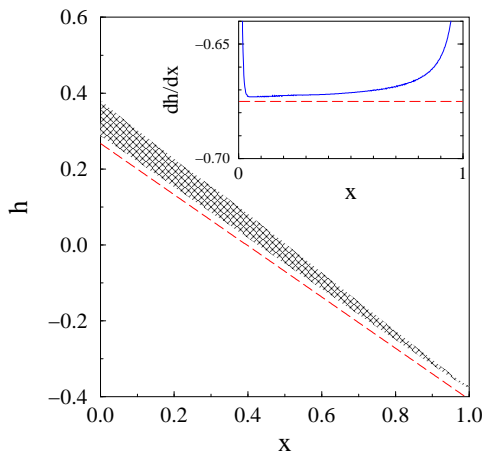,width=2.5in}
}

\caption{
The two-dimensional sandpile immediately after the failure zone has propagate
all the way to the top.  The static layer below the rolling layer follows
closely the critical slope (dashed line).  However, as shown in the inset, the
slope of the static layer (solid line) is slightly less steep than the critical
slope
(dashed line).
\label{FIG:CRITICAL1}
}

\end{figure}

\subsection{Two dimensional sandpile}

The numerical results indicate that the central subcritical
region is closely related to the metastable behavior of
the model. To understand the factors determining the limits
of metastability we consider the simpler case of a
two-dimensional sand-pile in a stationary container ($\Omega
= 0$ and $\alpha = 0$). This eliminates the complications of
the cylindrical geometry and the constant production of
flowing grains by the bucket rotation.

We start by preparing a static critical surface.  We first produce
a surface at the critical slope: $\partial h/\partial
x = -\tan \theta_f$ along with a small uniform rolling
layer.   Here $x$ is the linear position measured in units
of the system size $L_o$ with the high end of the pile at $x
= 0$ and the low end at $x = 1$. This constant slope surface is not
stationary due to the finite value of $\kappa$.  To obtain
the static critical surface we integrate Eq.\
(\ref{EQ:RHO}-\ref{EQ:H}) in one dimension until the 
initial flowing layer is completely converted to static. 

Once this static critical surface is obtained, we destabilize it by
tilting the surface through an
excess slope $\Delta S$.   We also assume the tilt produces
a uniform flowing layer $\Delta \rho$.   To determine the
characteristics of the resulting avalanche, we used a large $\Delta S =
0.1$ with the same parameters as we used for the spinning
bucket $\gamma = 500$, $D = 0.001$, $\kappa/\gamma =
1/500$ and $\Delta \rho = 10^{-8}$.  The mesh-size was
$\delta x = 0.001$ and time-step $\delta t = 2 \times 10^{-7}$. The
resulting avalanche has the same features as the avalanche
in the spinning bucket:

\begin{enumerate}

\item
The avalanche is induced by the growth of the rolling
layer at the foot of the pile.

\item
This buildup of the rolling layer creates a valley in the
static pile.  The large local slope of the static pile
causes the static material uphill to fail.

\item
The failure zone propagates uphill until it reaches the top
of the pile. As shown in Fig.\ \ref{FIG:CRITICAL1}, the
surface of the static pile below the rolling layer follows
closely the critical surface. However, a more detailed
examination (inset of Fig.\ \ref{FIG:CRITICAL1}) 
shows that the surface of the static
pile is slightly less steep than the critical slope.

\item
Since the propagation of the failure zone is faster than the convection
of grains down the hill, the total height of the pile (static + flowing)
is the same as the initial static pile.  Therefore, the
built-up flowing layer at $t^{*}$, the time the failure zone reaches
the top of the hill, is
$$
	\rho(x,t^{*}) \approx h(x,0) - h(x,t^{*}) = \Delta S ~ (1-x).
$$
This is confirmed in Fig.\ \ref{FIG:CRITICAL1} which shows a triangular
shaped flowing layer with $\rho(0,t^{*}) = \Delta S$ and $\rho(1,t^{*} = 0$.

\item
The rolling layer flows downhill leaving a subcritical
region at $x < 0.5$.  The surface for $x \gg 0.5$ follows
closely the critical surface except at the very edge where
there is a buildup of the static pile \cite{BOUTREUX}.

\end{enumerate}

\begin{figure}

\centerline{
\psfig{figure=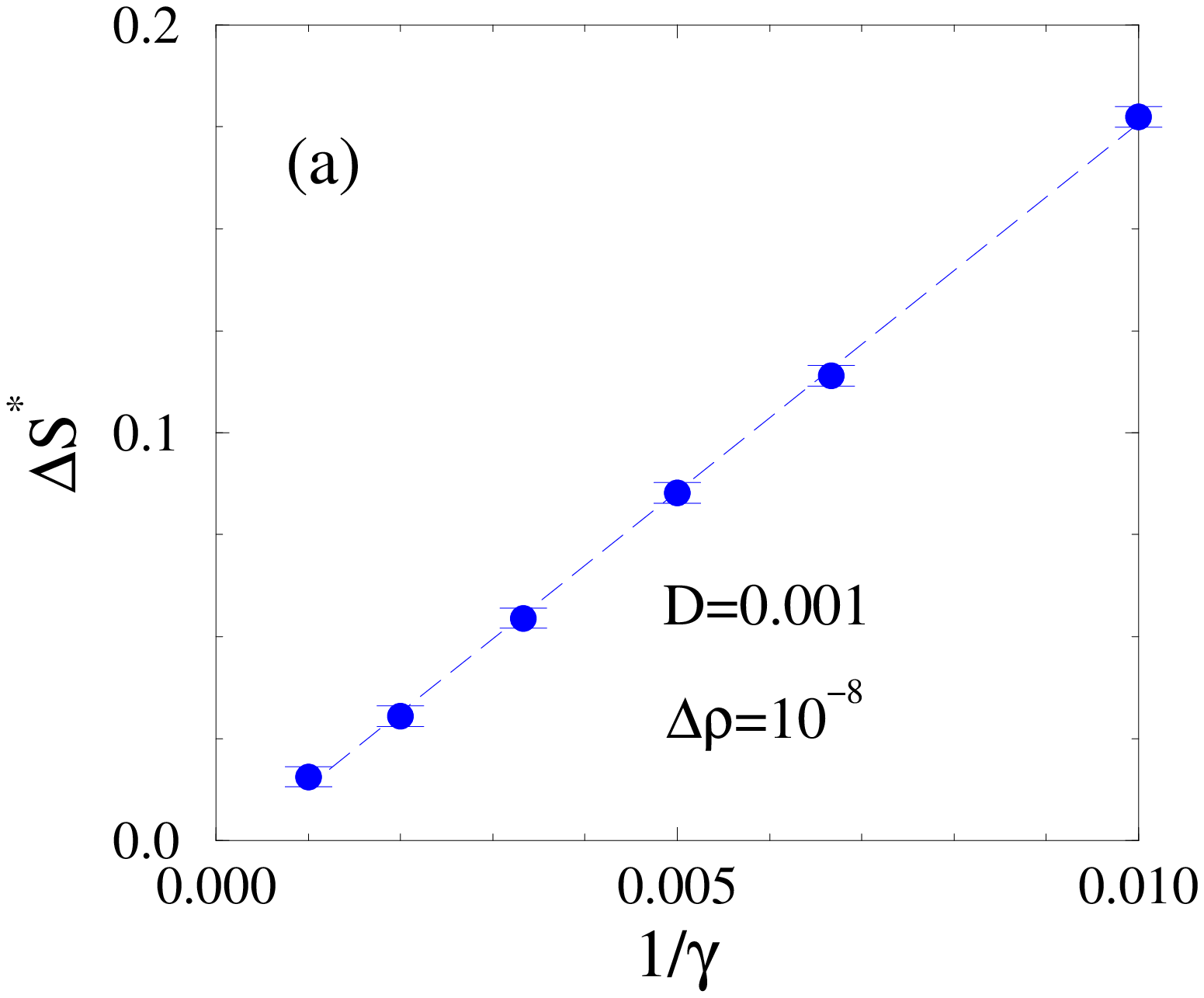,width=2.5in}
}

\centerline{
\psfig{figure=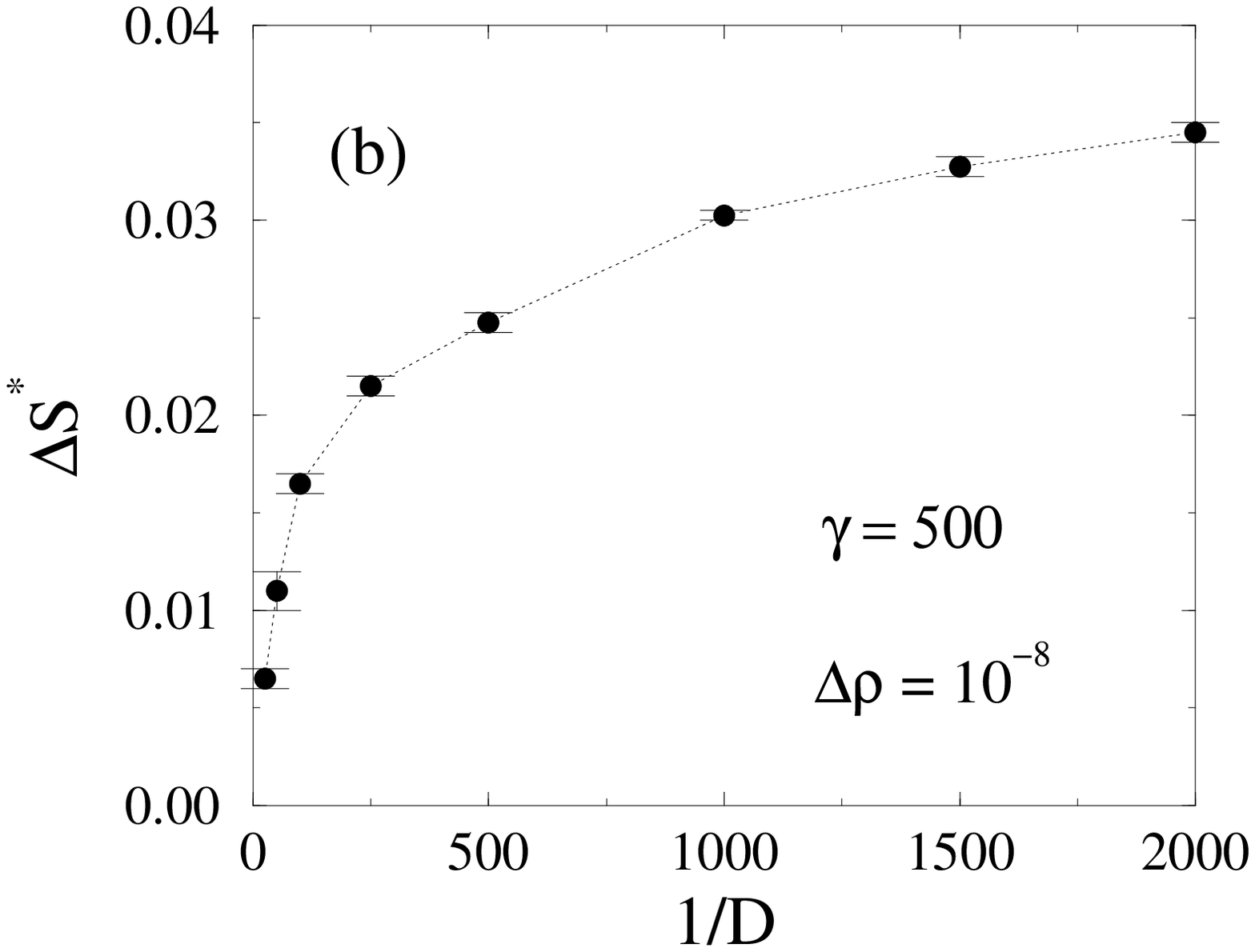,width=2.5in}
}

\centerline{
\psfig{figure=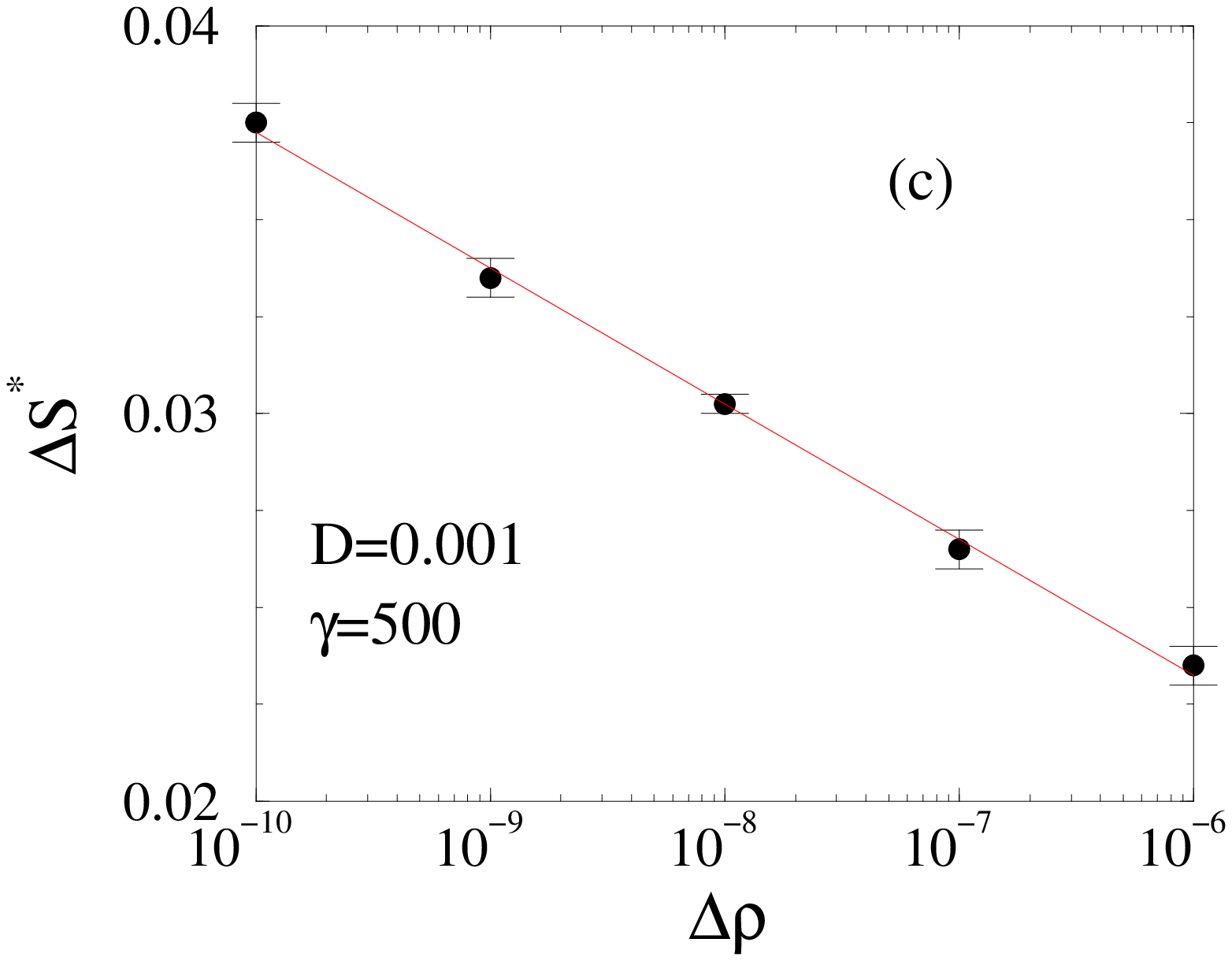,width=2.5in}
}

\caption{
(a)
The critical excess slope
$\Delta S^{*}$ as a function of $1/\gamma$.
The dashed line is
a straight line fit showing that $\Delta S \sim
1/\gamma$. (b)  $\Delta Sa^{*}$ as a function
of $1/D$.  For small $D$ (or large systems)
$\Delta S$ is only weakly dependent on $D$.
For large $D$, $\Delta S^{*} \sim 1/D$ indicating the linear
instability may initialize the avalanches in small systems.
(c)
$\Delta S^{*}$ as a
function of the initial rolling layer $\Delta \rho$.
The dashed line is a straight line fit
showing that $\Delta S \sim - \log( \Delta \rho )$.
\label{FIG:DEPENDENCE}
}

\end{figure}

Since the magnitude of the subcritical region after an avalanche depends on the
excess slope before, we need to explore the limits of metastability. The linear
stability analysis predicts that the surface is stable until the excess slope
exceeds $\Delta S^{*} = 1/(D \gamma)$ \cite{BOUCHAUD}.  However, several
characteristics of the avalanche indicates that a different mechanism may 
destabilize the surface. First, the instability always begins at the bottom of
the pile, indicating that the system-size is important.   Second, the avalanche
is generated when there is an increase in the local steepness of the static
pile.  The linear stability analysis  neglects the effect of the flowing layer
on the static pile.

To check if the linear stability dominates the dynamics,
we
determined $\Delta S^{*}$, the minimum excess slope required
to initialize an avalanche, as a function of the different
parameters. Fig.\ \ref{FIG:DEPENDENCE}a shows $\Delta S^{*}$
as a function of the amplification rate $\gamma$.  Although
both the linear stability prediction and the observed
$\Delta S^{*}$ are proportional to $1/\gamma$, the observed
values are a factor of 20 smaller  the linear prediction.

A more serious conflict is shown in Fig.\ \ref{FIG:DEPENDENCE}b where $\Delta
S^{*}$ is plotted as a function of $D$.  The numerical results clearly deviate
from the linear stability prediction  that $\Delta S^{*} \propto 1/D$ for small
$D$.  Hence, for large systems, i.e., small $D$, the linear stability mechanism
is not relevant to the surface evolution in this limit.  On the other hand, for
large $D$, $\Delta S^{*} \sim 1/D$ so the linear instability mechanism may be
relevant for small systems.  However, for large $D$, the subcritical region
disappears and the final state follows closely the critical surface.

Fig.\ \ref{FIG:DEPENDENCE}c shows
$\Delta S$ as a function of the rolling layer $\Delta
\rho$.  Here $\Delta S^{*} \propto - \log( \Delta \rho )$
so that the metastability of the initial granular surface
depends on the amplitude of the initial perturbation. In
particular, a small but non-infinitesimal rolling layer is
required to induce an avalanche indicating that the avalanche is
initiated
through a nonlinear instability mechanism.

\section{Theory and Analysis}

\subsection{Nonlinear instability mechanism}

To understand the metastable behavior observed numerically
two effects must be considered:
First, the flowing layer is amplified as it flows downhill, and second,
the amplification of the flowing layer changes the slope of the static 
pile.

To do so we modify an argument given by Bouchaud et al.\ for the amplification
of a bump of rolling grain as it flows \cite{BOUCHAUD}. Consider a critical
surface tilted through an  initial excess slope $\Delta S$ with an uniform
flowing layer $\Delta \rho$.  Neglecting any change in the static pile,  Eq.\
(\ref{EQ:RHO}) shows that the flowing layer will grow exponentially as it flows
downhill. For small $D$, the flowing layer will approximately maintain its
shape so
\begin{equation}
	\rho(x,t) \approx \left\{
		\begin{array}{ll}
			0 & \mbox{if $x < t$}, \\
			\Delta \rho ~ e^{\gamma \Delta S t} &
				\mbox{if $x > t$}.
		\end{array}
					\right.
			\label{EQ:RHOX}
\end{equation}
In terms of the dimensionless variables both the radius of the bucket
and the speed of the grains is unity.  Therefore
for $t > 1$ all flowing material reaches the bottom leaving
the origin static state intact.

This argument indicates that the granular surface is stable
for all $\Delta S$.   However, this is the case only if
the change in the static pile can be neglected. The change
in $h(x,t)$ for small $D$ is
\begin{eqnarray}
	\Delta h(x,t)  &\equiv & h(x,t) - h(x,0) = 
		-\int^{t}_{0} dt' ~ \Gamma( \{ h \}, \{ \rho \} )
	\nonumber \\ & = &
		\approx
	 - \int^{t}_{0} dt' ~ \gamma \Delta S ~ \rho(t').
\end{eqnarray}
Substituting Eq.\ (\ref{EQ:RHOX}) for $\rho(t')$
and integrating to $t > 1$ gives (since $\rho(x,t) = 0$ if $t > x$),
\begin{eqnarray}
	\Delta h(x, t) =  - \Delta \rho
			 \left( e^{\gamma \Delta S x} - 1 \right),
\end{eqnarray}
Therefore the flowing layer generates an additional excess
slope of
\begin{equation}
	\Delta S'(x) = -\der{ \Delta h }{ x } -
		=  \gamma \Delta S ~ \Delta \rho  ~ e^{\gamma \Delta S x }
	\label{EQ:EXCESS1}
\end{equation}
If $\Delta S'$ is not small there will be  a positive feedback mechanism. The
increased steepness makes the flowing layer grow faster, which in turn,
generates an even larger excess slope. Once this positive feedback mechanism
builds up, the increased slope causes the material uphill to fail and the
failure zone propagates up to the top of the hill. Since $\Delta S'(x)$ is
largest at $x = 1$ the avalanche must start at the bottom of the pile in
agreement with numerical observations.

Using Eq.\ (\ref{EQ:EXCESS1}) the minimum
initial excess angle required for an avalanche occurs when
$\Delta S'(1) \approx 1$ or
$\gamma \Delta S^{*} ~ \Delta \rho ~ e^{\gamma \Delta S^{*}}  \approx 1$.
Solving for $\Delta S^{*}$ gives
\begin{equation}
	\Delta S^{*}  \approx  -\frac{ 
		\log(  \Delta \rho ) + \log( \gamma \Delta S^{*} )
			}{ \gamma}
			\approx -\frac{ \log \Delta \rho }{ \gamma },
				\label{EQ:DELTASSTAR}
\end{equation}
where we assume $\Delta \rho \ll  1$.
This result is in agreement with our numerical results of the previous
section: $\Delta S^{*} \sim 1/\gamma$, $\Delta S^{*}
\sim -\log( \Delta \rho )$ and $\Delta S^{*}$ approximately
independent of $D$ for small $D$.

A more intuitive understanding is obtained by
rewriting the metastability criteria in terms of the original
unscaled parameters.  Using $\gamma = \gamma_o L_o/v_o$ and
$\Delta \rho = \Delta \rho_o/L_o$ to rewrite Eq.\  \ref{EQ:DELTASSTAR}
in terms of the dimensionful variables we have
\begin{equation}
	\Delta S^{*}
	\approx -\frac{ v_o  }{ \gamma_o L_o}
		\log\left( \frac{\Delta \rho_{o} }{L_o} \right).
\end{equation}
Therefore, according to this model, the range of metastable
slopes and hence, the Bagnold angle, depends on the system
size $L_o$.   Furthermore $\Delta S^{*}$ vanishes as $L_o \to
\infty$ showing that the metastability vanishes for infinite systems. 
This is because it takes time $t_o = L_o/v_o$ for the
flowing material at the top of the hill to reach the bottom.
During this time the flowing layer is amplified by a factor
$e^{\gamma_o \Delta S ~t_o}$.  The larger the system the more
the flowing layer is amplified and the more likely
an avalanche is initiated.

\subsection{Central subcritical region}

We now return to the origin of the central subcritical region observed
after an avalanche.  Assume  an avalanche is initiated
at the bottom of the pile.  For large $\gamma$  and small $D$, we
empirically find that
failure zone propagates uphill much faster than the rate at which grains
are convected downwards.  We find that
the
static pile below the flowing layer is slightly less steep than the
critical surface (see Fig.\ \ref{FIG:CRITICAL1}).
Since the slope of the static pile is less than critical, the
built-up rolling layer is continually converted to static as the
layer flows down the hill.  The flowing layer is convected away from
the top of the pile first so this conversion occurs longest at the
bottom,
leading to final surface
that is shallower than the critical surface.

This argument can be made more quantitative by
approximating the built up rolling layer as a constant $\rho(x,t_1) =
\Delta S$ where $t_1$ is the time the failure
zone reaches the top of the pile.  As a further approximation,
we also assume that the underlying static pile has a constant
slope slightly less steep than critical
$\partial h/\partial x = -\tan \theta_f + S_D$ where $S_D$ is the
deviation of the static pile from the critical slope.  
For small $D$ we can neglect the change in the shape of the flowing
layer as it is convected downhill:
\begin{equation} 
	\rho(x,t) = \left\{
		\begin{array}{ll}
			0 & \mbox{if $t-t_1 > x$}, \\
			\Delta S ~ e^{-\gamma \Delta S t} &
				\mbox{if $t-t_1 > x$}.
		\end{array}
					\right.
			\label{EQ:RHOX1}
\end{equation}	
The change in the static pile after the flowing layer has passed  is
\begin{eqnarray}
	\Delta h(x,t)  = - \int^{t}_{t_1} dt' ~ \Gamma( \{ h \}, \{ \rho \} )
		=  \int^{t}_{t_1} dt' ~ \gamma S_D ~\rho(x,t').
\end{eqnarray}
Substituting Eq.\ (\ref{EQ:RHOX1}) for $\rho(t')$ and 
integrating to $t-t_1 > 1$ gives the total change in $h$,
\begin{eqnarray}
	\Delta h(x) =  \Delta S \left( 1 - e^{-\gamma S_D x} \right).
\end{eqnarray}
Therefore the slope of the static pile after the avalanche is
\begin{eqnarray}
	\pder{ h }{x} & = &
		-\tan \theta_f + S_D + \pder{ \Delta h }{ x }
			\nonumber \\
		& = &
		-\tan \theta_f + S_D + \gamma S_D ~ \Delta S ~ e^{-\gamma S_D x}
			\nonumber \\
		& \approx & 
		-\tan \theta_f + \gamma S_D ~ \Delta S ~ e^{-\gamma S_D x},
\end{eqnarray}
where we neglect the second term since $\gamma$ is large.
This result is in agreement with our qualitative picture.
In particular, the deviation of the static pile from the
critical surface is largest at small
$x$.

\section{Summary}

We have applied a model of granular surface flow developed by  Bouchaud et al.\
\cite{BOUCHAUD} and Mehta et al.\ \cite{MEHTA} to the spinning bucket
experiments of Baxter \cite{VAVREK,BAXTER}. The model qualitatively reproduces
the central subcritical region observed in the experiment at low rotation
rates.   The subcritical region occurs when a metastable surface becomes
unstable via a {\em nonlinear} instability mechanism. The nonlinear instability
is due to the amplification of the rolling layer as it is convected downhill. 
This amplified layer causes the static pile underneath to become steeper which
in turn causes the flowing layer to become even larger.  This positive
feedback mechanism initiates the avalanches in
``large'' systems. 
We numerically determined the excess slope required to destabilize
the metastable surface as a function
of system parameters and showed that it agrees with our
analytic arguments.  In particular, the model predicts that the  excess slope,
and hence the Bagnold angle, depends on system size and vanishes in the limit
of large systems.   Lastly we showed how the nonlinear instability leads to
the central subcritical region via a conversion of rolling grains to static
as they roll downhill.

\acknowledgements

I thank G.W.\ Baxter for many useful discussions and
comments and for providing the experimental data in 
Fig.\ \ref{FIG:LOWW}.  I am grateful to Ron McCarty for providing some of the
computational resources.  This work was supported by Cottrell College
Science Grant CC3993 from
Research Corporation.


\begin{references}
\bibitem{BARKERNOTE}
For example, Barker and Mehta have argued that the rotating drum experiment
is dominated by local rearrangement of grains rather than macroscopic flow
\cite{BARKER}.
\bibitem{PECLET}
These limits were discussed in Ref.\ \cite{BOUCHAUD} in terms of
a Pecl\'{e}t number $(v_o R)/D_o$ was called the Pecl\'{e}t
number in Ref.\ \cite{BOUCHAUD}.
\end{references}
\end{document}